\documentclass[pra,twocolumn,showpacs,floatfix]{revtex4}

\usepackage{graphicx}

\begin{document}               

\def\be{\begin{equation}}
\def\ee{\end{equation}}
\def\ba{\begin{eqnarray}}
\def\ea{\end{eqnarray}}
\def\bas{\begin{eqnarray*}}
\def\eas{\end{eqnarray*}}


\title{Density-functional theory for fermions in the unitary regime}
\author{T.~Papenbrock}
\affiliation{Department of Physics and Astronomy, University of Tennessee,
Knoxville, TN~37996, USA}
\affiliation{Physics Division,
Oak Ridge National Laboratory, Oak Ridge, TN 37831, USA}
\date{\today}

\begin{abstract}
In the unitary regime, fermions interact strongly via two-body
potentials that exhibit a zero range and a (negative) infinite
scattering length. The energy density is proportional to the free
Fermi gas with a proportionality constant $\xi$. We use a simple
density functional parametrized by an effective mass and the universal
constant $\xi$, and employ Kohn-Sham density-functional theory to
obtain the parameters from fit to one exactly solvable two-body
problem. This yields $\xi=0.42$ and a rather large effective mass. Our
approach is checked by similar Kohn-Sham calculations for the exactly
solvable Calogero model.
\end{abstract}

\maketitle

Ultracold atomic Fermi gases have received considerable experimental
and theoretical interest since the achievement of Fermi degeneracy by
DeMarco and Jin \cite{DeM99}. One of the most interesting features is
the ability to tune the inter-particle interaction itself via a
Feshbach resonance, and to study the system as it evolves from the BCS
regime with weakly attractive interaction toward the point where the
interaction is strong enough to form di-atomic molecules, and the
system may undergo Bose-Einstein condensation (BEC). Very recently,
the BEC-BCS crossover has been subject of numerous experimental
\cite{OHa02,Reg03,Bou03,Zwi03,Reg04,Kin04,Bar04,Bou04,Gre05} and
theoretical studies \cite{Hol01,Men02,Bul03,Kim04,Bul05}.

The dividing point of this BCS-BEC crossover defines the unitary
regime and is of particular interest
\cite{Ho03,Bul05b,Sch05,Car03,Ast04,Hei01}. Here, the two-body system
exhibits a bound state at zero energy, and the two-body scattering
length diverges. For dilute systems, the inter-particle distance is
much larger than the range of the interaction and much smaller than
the scattering length. Thus, the inter-particle distance is the only
relevant length scale, and the energy must be proportional to that of
a free Fermi gas, 
\be
\label{Carlson}
E(N)= \xi E_{TF}(N)
\ee
the dimensionless proportionality constant being denoted as
$\xi$. Here, $E(N)$ is the energy of the fully paired $N$-fermion
system while $E_{TF}(N)$ is the Thomas-Fermi energy of $N$
noninteracting spin-$1/2$ fermions.  The constant $\xi$ is universal,
as it describes the physics of any dilute Fermi gas in the unitary
regime. Approximately, it also describes systems close to the unitary
regime, and can also be applied to dilute neutron gases as the
two-neutron system also exhibits a scattering length that is much
larger than the range of the nucleon-nucleon interaction. The exact
determination of the universal constant $\xi$ is thus an important
task.

Recently, this constant was reliably determined through Monte Carlo
simulations as $\xi\approx 0.44\pm 0.01$ by Carlson {\it et al.}
\cite{Car03}, and as $\xi=0.42\pm0.01$ by Astrakharchik {\it et al.}
\cite{Ast04}. So far, simpler approaches have failed to agree on the
value of $\xi$, and they deviate considerably from the Monte Carlo
results.  Heiselberg \cite{Hei01} obtained $\xi=0.326$, while Baker
\cite{Bak99} found $\xi=0.326$ and $\xi=0.568$ from different Pad{\'e}
approximations to Fermi gas expansions.  Engelbrecht {\it et al.}
\cite{Eng97} obtained $\xi=0.59$ in a calculation based on BCS theory.

It is the purpose of this paper to present a simple calculation that
determines the universal constant $\xi$. It is based on Kohn-Sham
density-functional theory (DFT) with a two parameter density
functional that is fit to one analytically known result. The resulting
value $\xi\approx 0.42$ is close to recent Monte Carlo results.  This
paper is organized as follows. First, we present the DFT for the Fermi
systems in the unitary regime. The density functional has a
particularly simple and constrained form in the unitary regime.
Second, we test and validate the density functional through
calculations for the exactly solvable Calogero model.

Carlson {\it et al.} \cite{Car03} performed quantum Monte Carlo
simulations for systems of $N$ fermions in the unitary regime, with
the number of fermions $N$ ranging from $N=10$ to approximately
$N\approx 40$. They found in particular that the relation
(\ref{Carlson}) holds with very good accuracy for all even number
systems.  This is a remarkable finding, since exact quantum mechanical
energies usually differ from the corresponding Thomas-Fermi energies
due to finite size effects and shell effects, both of which are
apparently very small for fermions in the unitary regime. This
suggests that the density functional from Thomas-Fermi theory
\be
\label{fun_tf}
{\cal E}_{TF}[\rho] = \xi {\hbar^2\over m} c \rho^{5/3}
\ee
with $c={3\over 10} (3\pi^2)^{2/3}$ is a good approximation of the 
exact density functional, and that corrections might easily be accounted
for via full-fledged Kohn-Sham DFT.

As a starting point, we thus consider Thomas-Fermi theory of
harmonically trapped fermions in the unitary regime. The density functional is
\be
\label{tf_ho}
{\cal E}_{TF}^{HO}[\rho]= {\cal E}_{TF}[\rho] + {1\over 2}m \omega^2 r^2 \rho.
\ee
This functional is minimized under the condition that the density is 
normalized to $N$ particles. This yields the Thomas-Fermi density
\be
\label{den_ho}
\rho_{TF}(r)={1\over 3\pi^2}\left({2\over\xi l^2}\right)^{3/2} 
\left((3N)^{1/3}\xi^{1/2}-{r^2\over 2 l^2}\right)^{3/2}
\ee
where $l=(\hbar/m\omega)^{1/2}$ is the oscillator length of the harmonic trap.
For the Thomas-Fermi energy we insert the density (\ref{den_ho})
into the functional (\ref{tf_ho}) and integrate over space. This yields
\be
\label{TF}
E_{TF} = {1\over 4}(3N)^{4/3}\xi^{1/2} \hbar\omega.
\ee

For the two-fermion system, the exact quantum mechanical result for the 
energy is \cite{Bus98}
\be
\label{E_ex}
E_{\rm ex}=2\hbar\omega. 
\ee
Let us assume that the relation (\ref{Carlson}) also holds for
harmonically trapped systems. Thus, we might equate Eq.~(\ref{E_ex})
with Eq.~(\ref{TF}) for $N=2$, and solve for the universal constant
$\xi$. This yields $\xi=64/6^{8/3}\approx 0.54$. Note that this simple
result deviates only about 20\% from the Monte Carlo results
\cite{Car03,Ast04}. This is quite encouraging and motivates us to
compute a more accurate estimate for $\xi$ via Kohn-Sham DFT.

In Kohn-Sham DFT \cite{Koh65}, the ground-state density and energy of an
interacting $N$-fermion system is obtained from varying the (generally
nonlocal and unknown) density functional. Unfortunately, there is 
no simple recipe that permits one to construct the density functional. 
For dilute systems with sufficiently
small range and positive scattering length, a systematic approach has been
given by Puglia {\it et al.} \cite{Pug03}. For electronic systems, one
usually parametrizes the density functional in terms of local
densities and their gradients, and fits the considerable number of
parameters to experimental data and theoretical results for infinite
systems. This elaborate and cumbersome approach has been successfully
implemented in recent decades in quantum chemistry (see, e.g.,
Ref.~\cite{Tao03} and references therein), and a similar approach is
also pursued in nuclear structure \cite{Gor02,Ben03,Lun03,Sto03}. Note
that DFT has also been successfully applied to study the BCS-BEC
cross-over. In their study, Kim and Zubarev \cite{Kim04} employed a
density functional that was fit to known results in the regime of very
small scattering length and to the Monte Carlo results for the unitary
regime, and employed a Pad{\'e} approximation for intermediate values
of the scattering length. 

Fortunately, the case of the unitary regime is much simpler. In what
follows we restrict ourselves to small even-number systems with equal number
of fermions in the two spin states. As the interaction does not
introduce any new length scale into the system, a local density
functional can only contain density terms proportional to
$\rho^{5/3}\hbar^2/m$, and gradients of the form $\rho^{-(2k-2)/3}
(\partial^{2k}\rho) \hbar^2/m$, with integer $k>0$. Here, we allow only 
for the density dependent term and incorporate gradient terms through an
effective mass. Our ansatz for the density functional thus becomes 
\be
\label{df}
{\cal E}[\rho] =
{\hbar^2\over m} \left[ {m\over 2 m_{\rm eff}}\sum_{j=1}^N
|\nabla \phi_j(\vec{r})|^2 + \left(\xi-{m\over m_{\rm eff}}\right) 
c \rho^{5/3}\right].
\ee
The density is given as $\rho=\sum_{j=1}^N |\phi_j(\vec{r})|^2$.  The
universal constant $\xi$ and the effective mass $m_{\rm eff}$ are
parameters that will be determined below.  The effective mass is, in
principle, $N$-dependent, but we suppress this dependency here. Note
that this density functional has two important properties: First, its
Thomas-Fermi limit is given in Eq.~(\ref{fun_tf}) and is thus
proportional to the density functional of the free Fermi gas. Second,
non-localities of the density are introduced through the effective
mass. Note that more elaborate approaches for superfluid systems also
introduce pairing densities \cite{Oli88,Bul02,Yu03,Per04} in the density
functional in order to approximate the (unknown) non-local functional
for superfluid systems. The quality of the results presented below,
however, suggests that this is not necessary for the purpose of this
study.

\begin{figure}[t]
\includegraphics[width=0.45\textwidth]{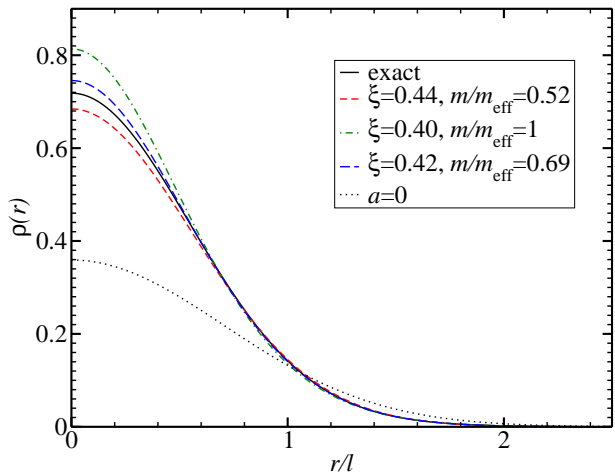}
\caption{\label{fig1}(color online) Density of the harmonically trapped
two-fermion system. Exact result (full line) compared to Kohn-Sham
results from three different density functionals (dashed lines and
dashed-dotted line), and the noninteracting system (dotted line).}
\end{figure}

For a determination of the parameters $\xi$ and $m_{\rm eff}$ we
consider the problem of two spin-1/2 fermions inside a harmonic trap
that interact via a zero-range interaction with infinite scattering
length. The ground state of the two-fermion system is a spin singlet,
and the relative coordinate wave function has been given by Busch
{\it et al.} \cite{Bus98}
\be
\label{psi_ex}
\psi_{\rm ex} (r) = (2\pi\lambda^3)^{-1/2}{\lambda\over r} 
e^{-\left( r / \lambda \right)^2/2}.
\ee
Here, $\lambda=\sqrt{2} l$ is given in terms of the oscillator length
$l$.  The wave function diverges like $1/r$ for small distances due to
the infinite scattering length; in practice, this divergence could be
cut off by any non-zero range of the interaction potential (see,
e.g. the discussion in Ref.~\cite{Car03}), and it does not cause any
problems in the normalization of the wave function. We recall that the
ground-state energy (\ref{E_ex}) of the two-fermion system is
considerably lower than for noninteracting fermions.  Employing the
relative wave function (\ref{psi_ex}) and the Gaussian ground state
for the center of mass coordinate, we arrive at the density
\be
\label{dens_ex}
\rho_{\rm ex}(r)={4\over \pi^{3/2} l^3} {l\over r} 
e^{-2\left(r / l\right)^2}
\int\limits_0^r dx \, e^{x^2}. 
\ee
A plot of this density is shown in Fig.~\ref{fig1} as a full line, and
can be compared to the noninteracting case (dotted line).

We add the term $(m/2)\omega^2r^2\rho(\vec{r})$ of the harmonic
confinement to the density functional (\ref{df}) and use Kohn-Sham
theory to compute the density for given sets of parameters $\xi$ and
$m_{\rm eff}$.  The best agreement with the exact density
(\ref{dens_ex}) and exact energy (\ref{E_ex}) is obtained for
$\xi=0.42$ and $m/m_{\rm eff}=0.69$. The resulting density is plotted
as the long-dashed line in Fig.~\ref{fig1} and exhibits only small
deviations from the exact density. The energy deviates about 0.1\%
from its exact value (\ref{E_ex}). Note that our value of the
universal constant is very close to the Monte Carlo results
\cite{Car03,Ast04}.

We want to estimate the robustness and stability of our result.  For
this purpose we consider the following two cases. First, we fixed the
value of the universal constant to the Monte Carlo result $\xi=0.44$
by Carlson {\it et al.} \cite{Car03}, and obtain the effective mass
$m/m_{\rm eff}=0.52$ from the best fit to the exact density and
energy. The resulting density is shown as the short-dashed line in
Fig.~\ref{fig1}, and the energy deviates less than 1\% from the exact
result. Second, we fix the effective mass to $m/m_{\rm eff}=1$, and
obtain $\xi=0.40$ from the fit of the density functional. The
resulting density is depicted as the dashed-dotted line in
Fig.~\ref{fig1}, while the energy again deviates less than 1\% from
the exact result. These results show that DFT yields quite robust
results for the universal constant $\xi$, and systematically improves
upon the naive Thomas-Fermi result.  This suggests that the good
agreement of our best value $\xi=0.42$ with the Monte Carlo results is
not merely accidental but due to the quality of the density functional
we employed.

We might even take our approach one step further and obtain a simple
analytical estimate for the universal constant $\xi$. Due to the
relatively large value of the effective mass, the prefactor of the
density-dependent term in the density functional (\ref{df}) becomes
rather small, and we might neglect it by simply setting $m/m_{\rm
eff}=\xi$. In presence of the confining harmonic potential, the
Kohn-Sham equation is then identical to the Schr\"odinger equation of
a three-dimensional harmonic oscillator where the kinetic energy is
modified by a factor $\xi$. The analytical result for the ground-state
energy of the two-fermion system is then $E=3\hbar\omega\xi^{1/2}$,
and comparison with the exact result (\ref{E_ex}) yields
$\xi=4/9$. Note that the resulting density is very close to the 
short-dashed line in Fig.~\ref{fig1} (labeled as $\xi=0.44,\, m/m_{\rm
eff}=0.52$).

In the unitary regime, the rather simple density functional (\ref{df})
does yield much improved and reliable results compared to Thomas-Fermi
theory. To further test the form of this density functional we
consider another interacting $N$-body system whose density functional
is also proportional to that of a free Fermi gas. The Calogero model
\cite{Cal69} with Hamiltonian
\ba
\label{CM} 
H &=& {\hbar^2 \over m}\sum_{j=1}^N
\left(-{1 \over 2}{\partial^2\over\partial x_j^2}
\,+\,\sum_{j<i} {{\beta\over 2}\left({\beta\over 2}-1\right)\over 
(x_i-x_j)^2}\right)
\nonumber\\
&& +\, {1\over2} m\omega^2\sum_{j=1}^N x_j^2,
\ea
is exactly solvable in one dimension. Here, $\beta\ge 1$ is a dimensionless 
coupling constant. 
The exact ground-state energy of this Hamiltonian is 
\be
\label{E_CM}
E_{\rm ex}=\hbar\omega \left( {N\over 2} + {\beta \over 4} N(N-1)\right).
\ee

Let us focus on the two-body interaction. The inverse square potential
does not introduce any new length scale into the Hamiltonian as it
scales as the kinetic energy. Thus, the inter-particle distance is the
only length scale, and the density-dependent energy must be
proportional to the one-dimensional Fermi gas, the proportionality
constant being denoted as $\eta^2$
\be
\label{CM_TF}
{\cal E}_{TF}[\rho] = \eta^2 {\pi^2\over 6}{\hbar^2\over m}\rho^{3}.
\ee
In this respect, the Calogero model is similar to the Fermi gas in the 
unitary regime. This approach leads directly to the Thomas-Fermi
theory for the Calogero model \cite{Sen95}. 

The constant $\eta^2$ can be determined from Thomas-Fermi theory once
we add the confining harmonic potential and thereby make contact with
exactly known results. Thus, the density functional becomes
\be
\label{d_tf}
{\cal E}[\rho]={\cal E}_{TF}[\rho] + {1\over 2} m \omega^2 x^2 \rho,
\ee 
and it is minimized by the normalized density
\be
\label{Wigner}
\rho_{TF}={1\over \pi \eta l}\sqrt{2N\eta - \left(r/l\right)^2}.
\ee
Here, $l=(\hbar/m\omega)^{1/2}$ again denotes the oscillator length. The
Thomas-Fermi density (\ref{Wigner}) is Wigner's semi circle, and
agrees with the standard approach \cite{Mehta}. 

We insert the density (\ref{Wigner}) into the functional (\ref{d_tf}) and 
integrate. The resulting Thomas-Fermi energy is 
\be
\label{E_cm_tf}
E_{TF}= {\eta \over 2}\hbar\omega  N^2, 
\ee 
and comparison with the exact result (\ref{E_CM}) fixes
$\eta=\beta/2$.  Note that the Thomas-Fermi energy (\ref{E_cm_tf})
differs from the exact result (\ref{E_CM}) by considerable 
finite size corrections, and this is a difference to the Fermi gas in the 
unitary regime.

Let us apply Kohn-Sham DFT to the Calogero model. The ansatz for the density
functional is in full analogy to the one we made for the Fermi gas in the 
unitary regime 
\ba
\label{CM_df}
{\cal E}[\rho]&=&{\hbar^2\over m} 
\left[ {m\over 2 m_{\rm eff}}\sum_{j=1}^N  
|\partial_x\phi_j(x)|^2 
+ {\pi^2\over 6} \left(\eta^2-{m\over m_{\rm eff}}\right)  \rho^{3}\right]
\nonumber\\
&&+{1\over 2}m\omega^2 x^2 \rho(x).
\ea
Here we have again introduced the effective mass $m_{\rm eff}$ as the   
only fit  parameter and have already added the term due to the harmonic 
confinement. 

\begin{figure}[b]
\includegraphics[width=0.45\textwidth]{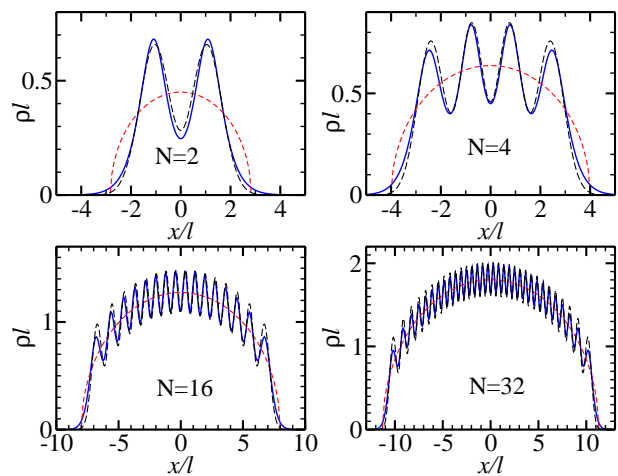}
\caption{\label{fig2}(color online) Densities of the Calogero model
for $\beta=4$ for different numbers of particles $N$. Full 
line: density functional theory; long-dashed  line: leading
order expansion of exact density; short-dashed line:
Thomas-Fermi theory.}
\end{figure}

We determine the effective mass by solving the Kohn-Sham equation for
the density functional (\ref{CM_df}) and compare the resulting density
with exact results. Note that the exact density of the Calogero model
(\ref{CM}) is only known for a few values of the coupling constant
$\beta$, though the many-body ground state of this model is
known for decades. For $\beta=1$, $\beta=2$, and $\beta=4$, the density is
related to the eigenvalue distribution of the orthogonal, unitary, and
symplectic Gaussian random matrix ensemble, respectively. Analytical
expressions are given in Refs.\cite{Kal02,Gar05}. We focus on the case
$\beta=4$, and determine the effective mass by fit. From calculations for 
particle numbers $N=2,4,16,32$ one obtains approximately
\be
\label{m_eff}
{m\over m_{\rm eff}}\approx 6.3 + {8.0\over N^2}. 
\ee 
Thus, the effective mass is considerably smaller than the
mass. Figure~\ref{fig2} shows that the Kohn-Sham densities are close
to the exact results. The Thomas-Fermi result is also shown for
comparison.  Note that only a relatively small effective mass
(\ref{m_eff}) reproduces the density oscillations. Note also that the
deviation of the DFT energies from the exact result is about half as
large as the error of the corresponding Thomas-Fermi energies.
This shows that the simple density functional (\ref{CM_df}) yields 
significantly improved energies and densities. 

In summary, we have used density-functional theory to compute the
universal constant of the Fermi gas in the unitary regime. This
approach is based on the observation that the Thomas-Fermi energy is a
reasonable first order approximation to the quantum mechanical
results, and on the constraints that the unitary regime imposes on the
form of the density functional. Our estimate $\xi=0.42$ results from a
best fit to the density and energy of the harmonically trapped
two-fermion system, and is in good agreement with much more elaborate
Monte Carlo studies. Our result is stable with respect to variations
of the density functional, and favors a sufficiently large effective
mass. The particular form of the density functional could also be
tested in applications to the Calogero model.

I thank J. Engel, P. Forrester, W. Nazarewicz and M. Stoitsov for
useful discussions. This research was supported in part by the
U.S. Department of Energy under Contract Nos.\ DE-FG02-96ER40963
(University of Tennessee) and DE-AC05-00OR22725 with UT-Battelle, LLC
(Oak Ridge National Laboratory).

\end{document}